\documentclass[sigconf, nonacm]{acmart}

\AtBeginDocument{%
  }
\usepackage{makecell}
\usepackage{multirow}
\usepackage{listings}
\usepackage[most]{tcolorbox} % 关键：加载 most 库（包含 multicol 和 breakable 支持）
\usepackage{lipsum} % 用于生成示例文本
\usepackage{color} % 用于生成示例文本
\usepackage{listings}
\usepackage{xcolor}
\usepackage{graphicx}
\usepackage{subcaption} 

\lstdefinestyle{mystyle}{
    backgroundcolor=\color{gray!10},   
    commentstyle=\color{black},
    keywordstyle=\color{blue},
    morekeywords={input,example},
    numberstyle=\tiny\color{gray},
    stringstyle=\color{red},
    basicstyle=\ttfamily\footnotesize,
    breakatwhitespace=false,         
    breaklines=true,                 
    captionpos=b,                    
    keepspaces=true,                 
    showspaces=false,                
    showstringspaces=false,
    showtabs=false,                  
    tabsize=2
}

\newtcolorbox{myquote}[1][]{
    colback=gray!10,
    colframe=gray!50,
    boxrule=0.5pt,
    arc=3pt,
    left=5pt,
    right=5pt,
    top=5pt,
    bottom=5pt,
    fontupper=\itshape\small,
    multicol,           % 允许跨栏
    breakable,          % 允许跨页
    #1
}

\newcommand\vldbdoi{XX.XX/XXX.XX}
\newcommand\vldbpages{XXX-XXX}
\newcommand\vldbvolume{14}
\newcommand\vldbissue{1}
\newcommand\vldbyear{2020}
\newcommand\vldbauthors{\authors}
\newcommand\vldbtitle{\shorttitle} 
\newcommand\vldbavailabilityurl{https://github.com/hnuGraph/LLM4DBdesign}
\newcommand\vldbpagestyle{plain}

\newtheorem{definition}{\bf Definition}%by zhwg
\newtheorem{problem statement}{\bf Problem Statement}%by zhwg
\newcommand{\inull}[1]{} 

\newcommand{\fullversion}[1]{}
\newcommand{\submitversion}[1]{}

\begin{document}
\title{Text2Schema: Filling the Gap in Designing Database Table Structures based on Natural Language}
% \title{Text2Schema: A New Paradigm for Data Management with Large Language Model}

%% College of Computer Science and Electronic Engineering, Hunan University
%% The "author" command and its associated commands are used to define the authors and their affiliations.

\author{Qin Wang, Youhuan Li}
\authornote{Corresponding author.}
\affiliation{%
  \institution{Hunan University}
  \city{Hunan}
  \country{China}
}
\email{{qinwang, liyouhuan}@hnu.edu.cn}

% \author{Youhuan Li}
% \affiliation{%
%   \institution{College of Computer Science and Electronic Engineering, Hunan University}
%   \city{Hunan}
%   \country{China}
% }
% \email{liyouhuan@hnu.edu.cn}

\author{Yansong Feng}
\affiliation{%
  \institution{Peking University}
  \city{Beijing}
  \country{China}
}
\email{fengyansong@pku.edu.cn}

\author{Si Chen, Ziming Li, Pan Zhang}
\affiliation{%
 \institution{Hunan University}
  \city{Hunan}
  \country{China}
} 
\email{{sichen, zimingli, hnuzhangpan}@hnu.edu.cn}

\author{Zihui Si, Yixuan Chen}
\affiliation{%
   \institution{Hunan University}
  \city{Hunan}
  \country{China}
}
\email{{szh-nine, cyx1218}@hnu.edu.cn}

\author{Zhichao Shi, Zebin Huang}
\affiliation{%
  \institution{Hunan University}
  \city{Hunan}
  \country{China}
}
\email{{shizhichao, hzb1031}@hnu.edu.cn}

\author{Guo Chen, Wenqiang Jin}
\affiliation{
   \institution{Hunan University}
  \city{Hunan}
  \country{China}
}
\email{{guochen, wqjin}@hnu.edu.cn}

% \author{
%   Qin Wang$^{1}$, Youhuan Li$^{1*}$, Yansong Feng$^{2}$, 
%   Chen Si$^{1}$, Ziming Li$^{1}$, Pan Zhang$^{1}$, \\
%   Zhichao Shi$^{1}$, Yuequn Dou$^{1}$, Chuchu Gao$^{1}$, 
%   Zebin Huang$^{1}$, Zihui Si$^{1}$, Yixuan Chen$^{1}$
% }
% \affiliation{
%   $^1$College of Computer Science and Electronic Engineering, Hunan University, China \\
%   $^2$Institute for Artificial Intelligence, Peking University, Beijing, China
% }
% \email{
%   {qinwang, liyouhuan}@hnu.edu.cn,
%   fengyansong@pku.edu.cn
% }

\begin{abstract}
People without a database background usually rely on file systems or tools such as Excel for data management, which often lead to redundancy and data inconsistency. Relational databases possess strong data management capabilities, but require a high level of professional expertise from users. 
Although there are already many works on Text2SQL to automate the translation of natural language into SQL queries for data manipulation, all of them presuppose that the database schema is pre-designed. In practice, schema design itself demands domain expertise, and research on directly generating schemas from textual requirements remains unexplored. 
In this paper, we systematically define a new problem, called Text2Schema, to convert a natural language text requirement into a relational database schema. 
With an effective Text2Schema technique, users can effortlessly create database table structures using natural language, and subsequently leverage existing Text2SQL techniques to perform data manipulations, which significantly narrows the gap between non-technical personnel and highly efficient, versatile relational database systems.
We propose \textit{SchemaAgent}, an LLM-based multi-agent framework for Text2Schema. We emulate the workflow of manual schema design by assigning specialized roles to agents and enabling effective collaboration to refine their respective subtasks. We also incorporate dedicated roles for reflection and inspection, along with an innovative error detection and correction mechanism to identify and rectify issues across various phases. Moreover, we build and open source a benchmark containing $381$ pairs of requirement description and schema. Experimental results demonstrate the superiority of our approach over comparative work.
\end{abstract}

\maketitle

\pagestyle{\vldbpagestyle}
\begingroup\small\noindent\raggedright\textbf{PVLDB Reference Format:}\\
\vldbauthors. \vldbtitle. PVLDB, \vldbvolume(\vldbissue): \vldbpages, \vldbyear.\\
\href{https://doi.org/\vldbdoi}{doi:\vldbdoi}
\endgroup
\begingroup
\renewcommand\thefootnote{}\footnote{\noindent
This work is licensed under the Creative Commons BY-NC-ND 4.0 International License. Visit \url{https://creativecommons.org/licenses/by-nc-nd/4.0/} to view a copy of this license. For any use beyond those covered by this license, obtain permission by emailing \href{mailto:info@vldb.org}{info@vldb.org}. Copyright is held by the owner/author(s). Publication rights licensed to the VLDB Endowment. \\
\raggedright Proceedings of the VLDB Endowment, Vol. \vldbvolume, No. \vldbissue\ %
ISSN 2150-8097. \\
\href{https://doi.org/\vldbdoi}{doi:\vldbdoi} \\
}\addtocounter{footnote}{-1}\endgroup

\ifdefempty{\vldbavailabilityurl}{}{
\vspace{.3cm}
\begingroup\small\noindent\raggedright\textbf{PVLDB Artifact Availability:}\\
The source code, data, and/or other artifacts have been made available at \url{\vldbavailabilityurl}.
\endgroup
}

\section{Introduction}
\begin{figure}[!bhtp]
	\centering
	\includegraphics[width=\linewidth]{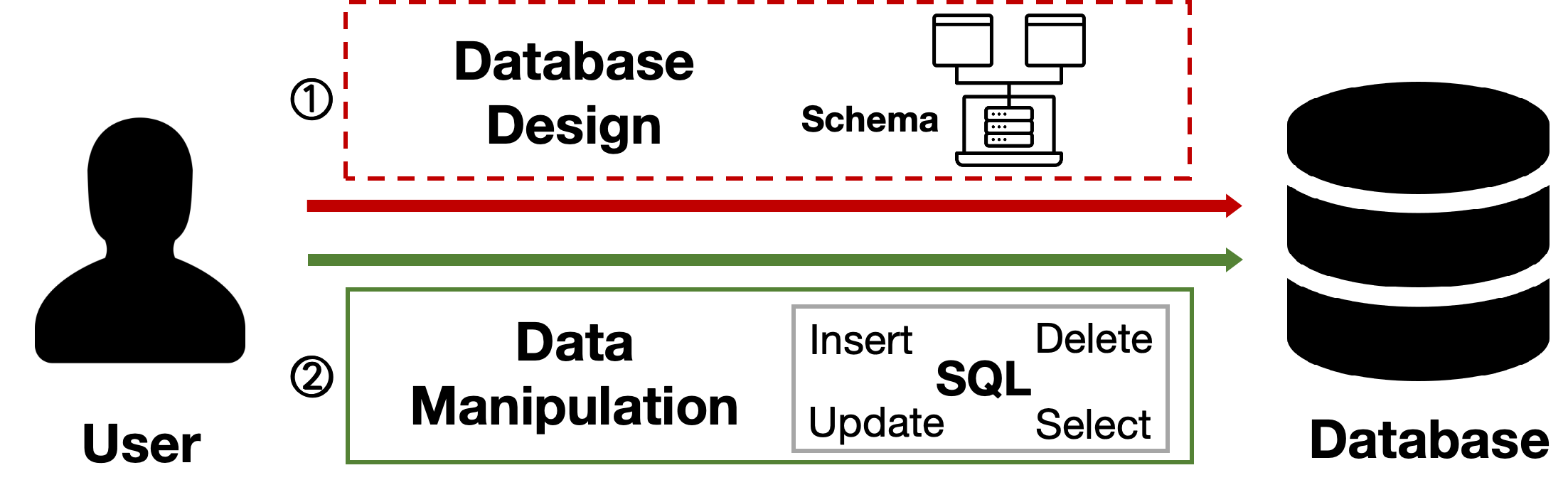}
	\caption {Interactions between ordinary users and databases.} 
\label{pic:user_database}
\end{figure}

Millions of users around the world rely on spreadsheet software for daily data processing. However, when faced with large datasets or complex requirements, such software is prone to data redundancy and inconsistencies. Although relational databases offer powerful data management capabilities~\cite{codd1970relational, codd1989relational, jatana2012survey}, they are largely inaccessible to most people due to a steep learning curve. 
As illustrated in Figure~\ref{pic:user_database}, enabling user-database interaction typically involves two steps: engineers (Step \textcircled{1}) design the schema to determine the table structure ; and then (Step \textcircled{2}) develop a simplified UI to map frequent data operations into fixed SQL statements, enabling ordinary users to interact with the database.
However, this model incurs significant costs and suffers from evident limitations due to its inflexible operational constraints.

Currently, numerous studies on Text2SQL (or NL2SQL)~\cite{xie2024mag, askari2024magic, xie2025opensearch} aim to convert data extraction requirements expressed in natural language into correctly executable SQL queries. They usually retrieve a set of candidate data columns from the database schema and then organize these columns into a correct SQL query statement under the SQL syntax, corresponding to Step \textcircled{2} in Figure~\ref{pic:user_database}.
However, these works all imply an underlying assumption: the database schema is already designed and readily available for data manipulations. This leaves a significant problem unaddressed. For individuals without expertise, the process of designing a robust database structure (Step \textcircled{1}) is still prohibitively difficult, often leading to poorly structured databases and data redundancy and inconsistency issues. To the best of our knowledge, there is no research that automatically generates a database schema directly from the natural language description of data requirements. Filling this research gap would be instrumental in empowering non-technical users with powerful database systems.

In this work, we propose a new problem named \textit{Text2Schema}, which focuses on directly generating a database schema from user requirements. We are the first to formally define a database schema, including table structures, primary-foreign keys, and domain constraints. 
Note that the database schema usually indicates the logical model data~\cite{silberschatz2011database}, which is different from the data definition language (DDL) statements that are specific to certain database products. In fact, it is easy to convert a database schema into DDL statements with auxiliary tools~\cite{Brdjanin2022} once a certain database system is determined. We also present a case study that demonstrates the conversion of a schema to a set of SQLite DDL statements (Section~\ref{sec:ddl_generation}).

Previous works related to this new problem could be divided into two categories. The first is Text2SQL that converts natural language into executable SQL queries~\cite{liu2025survey}. These works do not generate database schemas and could not solve the proposed problem. The second is automated database design~\cite{ruoff1984codes, lloyd1992expert, thonggoom2011semi, wand2017thirty, brdjanin2022towards}. 
All of them focus on the automation of conceptual design with customized rules or traditional deep learning models, outputting entities and relations. These works neither generate data types, constraints, nor conduct normalization that are important for the logical schema.
Another potential approach is to directly apply large language models (LLMs)~\cite{achiam2023gpt, touvron2023llama, team2023gemini, yang2024qwen2} to generate schemas due to their strong reasoning abilities.
However, we find that this straightforward approach is poor in effectiveness.

\begin{figure}[!t]
	\centering
	\includegraphics[width=3.3in]{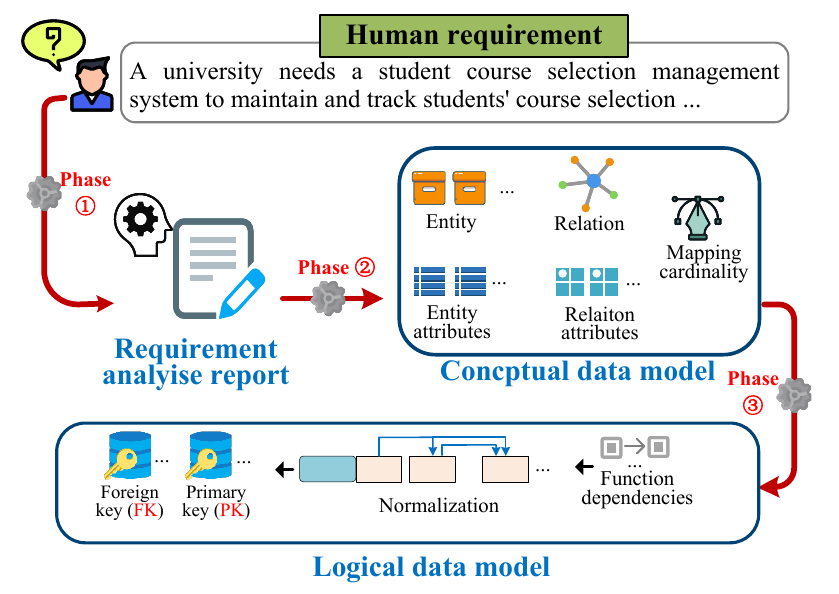}
	\caption {The process for database schema design. The 1st phase is requirements analysis. The 2nd is conceptual design phase for an Entity-Relationship (ER) model. The 3rd is logical design phase mapping ER model into logical schema.} 
\label{pic:workflow}
\end{figure}

To address this, we propose an LLMs-based multi-agent framework, called \textit{SchemaAgent}, to automatically generate schemas that satisfy the third normal form (3NF)~\cite{codd1972further}, as a 3NF schema is sufficient in most cases~\cite{3nf-enough-ma2013application}. 
% 这里逻辑有点不对，修改
Specifically, database schema design mainly consists of three phases, that is, (1) user requirement analysis; (2) conceptual design; (3) logical design (schema)~\cite{khennoufa1994towards, helskyaho2015oracle}. We first assign an agent to each of the three subtasks in schema generation (see Figure~\ref{pic:workflow}): Product manager for requirement analysis, Conceptual data model designer, and Logical data model designer. 
However, we find that error rates tend to be high in conceptual data modeling due to the difficulty in determining the appropriate sets of entities, relations, and mapping cardinality.
It could have a substantial impact on the quality of the schema if these errors are not corrected before we map the conceptual model into tables, columns, or constraints.
Hence, we introduce the $4$-th role: a reviewer specifically to supervise and validate the conceptual model in a timely manner.
This reviewer significantly reduces errors in the conceptual model.
Additionally, to further validate the completeness and correctness of logical model against user requirements, we design a pair of QA engineer and Test executor in SchemaAgent, where the QA engineer produces test cases based on the requirement while the Test executor conducts the corresponding tests to evaluate the schema quality.

The sequential workflow of these roles may suffer from the accumulation of errors, as we find that errors are often discovered later rather than in a timely manner.
Therefore, we design a group chat~\cite{wu2023autogen, hong2023metagpt} based communication mechanism to reduce error accumulation, where agents in the workflow could assist in identifying the errors made by previous ones.
By pinpointing the exact phase where the error occurred and providing timely feedback, our intersection mechanism guides the relevant roles to refine the solution. This dynamic feedback loop improves both the accuracy and efficiency of the schema generation process.

Furthermore, we develop the first specialized relational database schema benchmark~\textit{RSchema}, which contains $381$ pairs of requirement and the corresponding schema covering various real-world scenarios. 
Due to the complexity of schema generation, it is time-consuming and laborious to create Rschema, which has undergone multiple rounds of refined construction by database groups.
Extensive experiments confirm that our method outperforms both Chain-of-Thought (CoT) and direct prompting approaches across GPT-3.5-turbo, GPT-4o, and DeepSeek-v3. 

Our contributions are summarized as follows:
 \begin{itemize}
     \item We are the first to propose Text2Schema and fill a significant gap in converting natural language requirements into database schemas. We also provide the first formal definition of a database schema. (Section~\ref{subsec:task_definition})
     \item We are the first to propose an LLM-based multi-agent framework for Text2Schema (Section~\ref{sec:preliminary}). This novel problem will also give rise to a series of interesting future research directions (Section~\ref{sec:future_work}).
     \item We design six roles in our framework and propose a controllable error detection and correction mechanism to significantly reduce accumulated errors, guaranteeing the schema quality (Section~\ref{Method}).
     \item We create the first database schema generation benchmark, including $381$ schemas in diverse scenarios, as well as automatic evaluation metrics (Section~\ref{Corpus Construction}).
     \item The experimental results show that our framework significantly outperforms comparative methods (Section~\ref{Experimental Results and Analysis}).
 \end{itemize}

\section{Preliminary} \label{sec:preliminary}
In this section, we would discuss database schema (Section~\ref{subsec:schema-discuss}), and then give it a first formal definition (Section~\ref{subsec:task_definition}). We would also discuss the literature and distinguish our work from previous work from both the problem and approach perspectives (Section~\ref{subsec:relatedwork}).

\subsection{Schema of Different Level} \label{subsec:schema-discuss}
We primarily focus on the logical schema, as it is the default database schema by design~\cite{silberschatz2011database}.
There are three types of schema in database design: conceptual schema (from conceptual data modeling), logical schema (from logical data modeling), and physical schema (from physical design).
The conceptual schema lacks important constraints, such as foreign key constraints and domain constraints. Additionally, it has not undergone relational normalization and fails to meet the requirements of the essential third normal form (3NF).
The logical schema depicts the table structures, constraints (including data types).
It usually requires normalization to be in 3NF.
It is the most important one, which is generally used by programmers to construct applications~\cite{silberschatz2011database}. 
The physical schema is hidden beneath the logical one and can generally be easily changed without affecting application programs~\cite{silberschatz2011database}.
This paper focuses on the implementation of the logical schema (hereafter referred to as "schema"). Our future research would incorporate physical design, encompassing denormalization for frequent queries and the establishment of indexes to enhance query performance.

\subsection{Problem Definition} \label{subsec:task_definition}
A logical schema is a structured collection of components that defines the logical organization and constraints of data within a database system. 
To the best of our knowledge, no prior studies have provided a formal definition of the database schema. 
We formally define the database schema in the following Definition~\ref{def:schema}.

\newcommand{\mcs}{\mathcal{S}}
\newcommand{\mcr}{\mathcal{R}}
\newcommand{\mca}{\mathcal{A}}
\newcommand{\mcp}{\mathcal{P}}
\newcommand{\mcf}{\mathcal{F}}
\newcommand{\mcd}{\mathcal{D}}

\begin{definition}[Database Schema] \label{def:schema}
A database (logical) schema, denoted as $\mcs$, is a 5-tuple:
$\mcs = \{\mcr, \mca, \mcp, \mcf, \mcd\}$, where 
\begin{itemize}
    \item 
    $\mcr$ is a set of identifiers of relations (tables) within $\mcs$.
    For each $r_i\in \mcr$, $r_i$ is essentially a pair $\langle tID, tName \rangle$ corresponding to a table, where $tID$ and $tName$ are the corresponding table ID and table name, respectively;
    \item 
    $\mca$ is a set of attributes (columns), where each $a_i\in \mca$ contains attribute ID (denoted as $aID$), name ($aName$), data type ($aType$), and $tID$ indicating the table to which $a_i$ belongs; we may use $\mca(tID)$ to indicate the set all attributes belonging to the table of ID $tID$;
    \item 
    $\mcp$ denotes the set of primary key constraints. Each element $p_i$ in $\mcp$ exactly contains the primary keys of a unique table of ID $tID$. And, naturally, $p_i$ is a non-empty subset of $\mca(tID)$;
    \item 
    $\mcf$ denotes foreign key constraints, and each $f_i\in \mcf$ is a pair of attributes $\langle a_{i_1}, a_{i_2} \rangle$ indicating that foreign key attribute $a_{i_1}$ references the primary key attribute $a_{i_2}$;
    \item 
    finally, $\mcd$ is the set of domain constraints, and each $d_i \in \mcd$ is a pair $\langle a_i, c_i \rangle$ where $a_i$ is an attribute while $c_i$ indicates one of the domain constraints in $\{\text{not null}, \text{unique}\}$.
\end{itemize} 
\end{definition}

This is the first formal definition of database schema, and we omit user-defined constraints that do not have a specific or fixed form, since it could involve an indefinite number of tables or attributes.

\begin{definition}[Text2Schema]
We propose to study the problem of converting text descriptions of business requirements into a database schema. We use Text2Schema to denote this new problem, and we require the output schema to be in 3NF.
\end{definition}

\subsection{Related Work} \label{subsec:relatedwork}
\subsubsection{Relational Database Design}
Since the advent of relational databases, their design has remained a popular research. A multitude of methodologies has been developed to tackle specific aspects of relational database design.

Conceptual modeling is acknowledged as the most pivotal stage in the database design process~\cite{thalheim2009extended}. Among the various approaches, text-based methods remain the most traditional and foundational. These methods can be categorized into several distinct types: linguistics-based (e.g., LIDA ~\cite{overmyer2001conceptual} and ER-converter~\cite{omar2004heuristic}), pattern-based (e.g., APSARA~\cite{purao1998apsara}), case-based (e.g., CABSYDD~\cite{choobineh2004cabsydd}), and ontology-based (e.g., OMDDE~\cite{sugumaran2002ontologies} and HBT~\cite{thonggoom2011semi}).
Moreover, Textodata~\cite{brdjanin2022towards} offers an automated solution for converting natural language text into a target conceptual database model represented by a UML class diagram. Despite these advancements, existing methods often lack deep semantic understanding, resulting in poor performance when handling complex scenarios. 

During the logical design phase, conceptual models are systematically transformed into relational models. The foundational approach to this transformation is first introduced by~\cite{teorey1986logical}. Subsequently, \cite{borgida2009logical} introduced improved strategies for mapping Enhanced Entity-Relationship (EER) diagrams to relational models.

The functional dependency (FD) discovery is a fundamental step in the normalization of logical database schemas, as normalization requires identifying all FDs among the attributes. FD discovery has attracted sustained research interest within the data management community. ~\cite{huhtala1999tane} introduced the TANE algorithm, which efficiently discovers FDs from large databases using several pruning strategies to reduce the search space.~\cite{liu2013discovery} proposed a hash-based method that leverages hash tables to check FD satisfaction.~\cite{wyss2001fastfds} developed the FastFDs algorithm, which performs FD discovery in a depth-first and heuristic-driven manner over a search tree.~\cite{papenbrock2016hybrid} proposed a hybrid method, HYFD, which can handle datasets with many tuples and columns. Based on this,~\cite{wei2019discovery} designed DHyFD, a dynamic hybrid algorithm that further improves the performance of HYFD. Additionally,~\cite{wan2023efficient} proposed FSC algorithms, which are tailored for FD discovery in large-scale datasets. 
All of the aforementioned methods rely on analyzing existing datasets to extract functional dependencies. In contrast, benefiting from the rich knowledge embedded in LLMs, this paper aims to automatically discover FDs during database design, thereby enabling normalization of the logical schema. 
% To the best of our knowledge, this work presents the first end-to-end pipeline that transitions directly from the user requirement to a normalized logical schema design.

\begin{figure*}[!hptb]
	\centering
	\includegraphics[width=\textwidth]{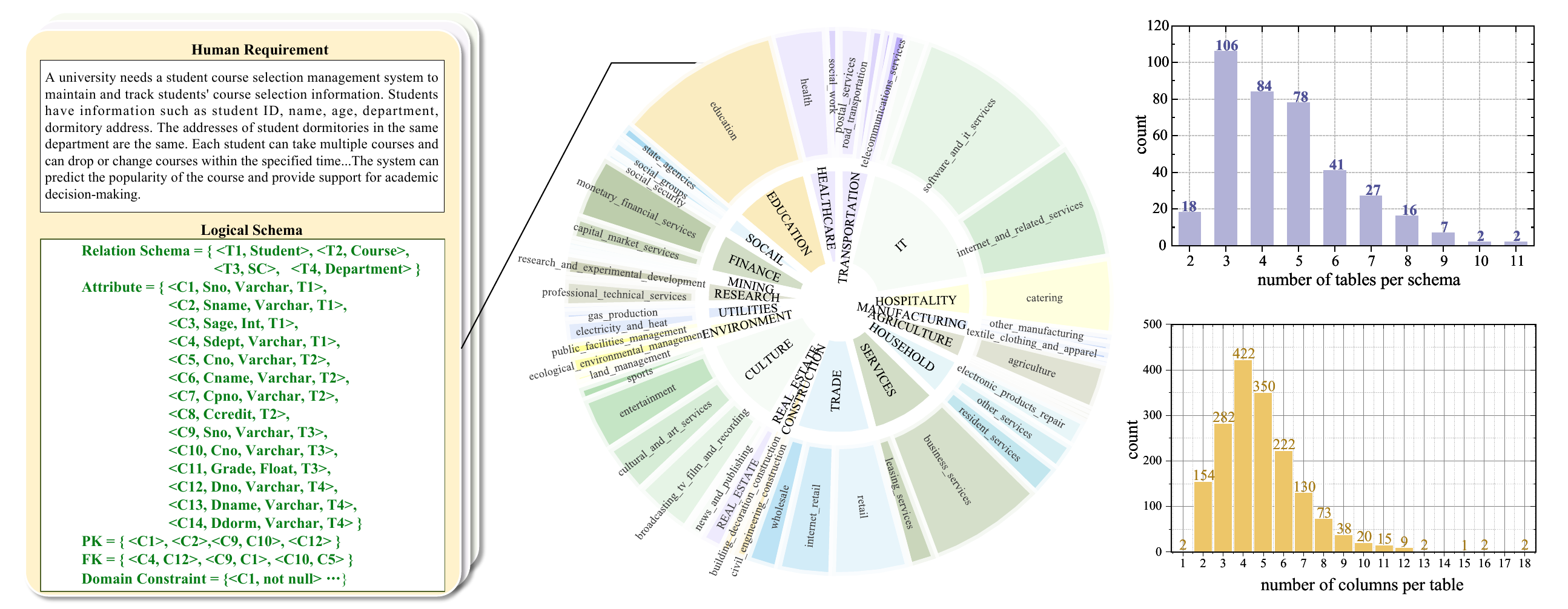}
        \caption{Structural overview of our RSchema database. The central area illustrates the distribution of domains. The left panel presents an example schema derived from a specific domain. The upper right panel depicts the distribution of the number of tables per schema, while the lower right panel shows the distribution of the number of columns per table, providing insights into the schema complexity and granularity.} 
\label{pic:schema_distribution}
\end{figure*}

\subsubsection{LLM-based multi-agents applications}
Single-agent systems that utilize LLMs have achieved notable advances through techniques such as problem decomposition~\cite{khot2022decomposed}, tool utilization~\cite{zhao2024let, li2023api}, and memory storage~\cite{bai2024transformers} during environmental interactions. 
Building on these developments, multi-agent systems have further expanded the capabilities of LLMs by specializing them into task-specific agents and enabling collaborative decision-making among autonomous agents. 
Recent research highlights the success of multi-agent systems in areas such as software development~\cite{hong2023metagpt, wu2023autogen, he2025llm} and biomedical, financial, and psychological domains. 
In the domain of databases, multi-agent systems have been applied primarily to tasks such as Text-to-SQL~\cite{xie2024mag, askari2024magic, xie2025opensearch}, query optimization~\cite{wang2024tool}, and database diagnostics~\cite{zhou2023d}. These applications demonstrate the potential of multi-agent systems to enhance the efficiency, accuracy, and adaptability of database-related processes.

\section{Corpus Construction}
\label{Corpus Construction}
We develop the dataset \textit{RSchema} containing $381$ samples covering various domains. Each sample is a pair of requirement text and the corresponding logical schema. 
Neither academia nor industry has publicly available database requirements and the corresponding schemas due to privacy concerns.
To the best of our knowledge, this is the first schema generation benchmark.
The entire construction process takes two months, consisting of four phases, where the first is the collection of raw data and the generation of the primary sample (Section~\ref{data_collection}), while the second and third are two rounds of sample refinement by eleven members of a database group (Sections~\ref{schema_annotation} and \ref{schema_review}).
The final phase is to review all samples by an experienced expert and a selected annotator (Section~\ref{final_review}).
Figure~\ref{pic:schema_distribution} illustrates the distribution of our dataset. In total, the dataset comprises 19 major categories.
Statistical analysis reveals that the average user requirement text contains 164 words, which is sufficient to express the core user requirements.

\subsection{Primary Sample Generation}
\label{data_collection}  
We create more than $500$ primary samples that would be refined later by experts.
These primary samples are generated on the basis of raw data that are collected from three distinct sources.
The first source is the SchemaPile dataset~\cite{dohmen2024schemapile} of more than $20,000$ DDL (Data Definition Language) files, and we transform these DDL contents into nearly $136$ primary schemas.
We employ an LLM Qwen2.5-72B-Instruction~\cite{yang2024qwen2} to generate the corresponding primary requirement descriptions for these DDL files. 
Secondly, inspired by the data generation capabilities of LLMs \cite{tang2023does}, we craft nearly $93$ primary requirement descriptions and schema pairs that span various industry scenarios. 
Lastly, we also construct nearly $289$ primary samples on materials that are crawled from the Internet, such as graduate theses, database design exams, Entity-Relationship (ER) diagrams, and technical blogs.

\subsection{Refinement Annotation}
\label{schema_annotation}
Due to the inherent noise in the raw data, manual annotation is crucial to ensure data quality.
During the annotation process, we first refine the requirement text to ensure its reasonableness.
Next, we follow the rigorous database design process to obtain a schema that is consistent with the requirements.
Specifically, we identify components such as entity sets, relationship sets, and attributes to systematically construct the conceptual model. Then, we conduct dependency-preserving decomposition and identify the keys to build the normalized logical schema.
Annotators would conduct detailed analysis obeying design principles, to guarantee the quality of these samples. We employ four key quality assurance measures: (1) each requirement should correspond to an explicit scenario and (2) each schema should support possible operations according to the corresponding requirement; (3) each schema must satisfy the third normal form (3NF), which is a convention~\cite{3nf-enough-ma2013application}; (4) the average length of requirements assigned to each annotator is well balanced.
Each annotation takes $25$ minutes on average.

\subsection{Cross Review}
\label{schema_review}
When a sample is annotated, the accuracy of the corresponding schema would be verified by another annotator to ensure the quality of the requirement text.
In this way, disagreements may occur among different annotators for the same sample.
We resolve this by setting additional discussions, and the sample would be repeatedly refined until the annotators reach consensus.
We also record the discussions (opinions of annotators) as a reference for the final phase.
Each review takes $15$ minutes on average.

\subsection{Final Review}
\label{final_review}
The final review is conducted by an experienced expert and a selected annotator.
A sample would be considered reliable if both the expert and the annotator confirm its quality.
If there is a disagreement, the experienced expert will make the final decision. 
In addition, a sample would be deleted if neither the expert nor the annotator accepts it.
This meticulous process costs one week and ultimately produces $381$ samples.

\section{Method}
\label{Method}
\subsection{Agent Structure}
\begin{figure*}[!t]
	\centering
	\includegraphics[width=6.0in]{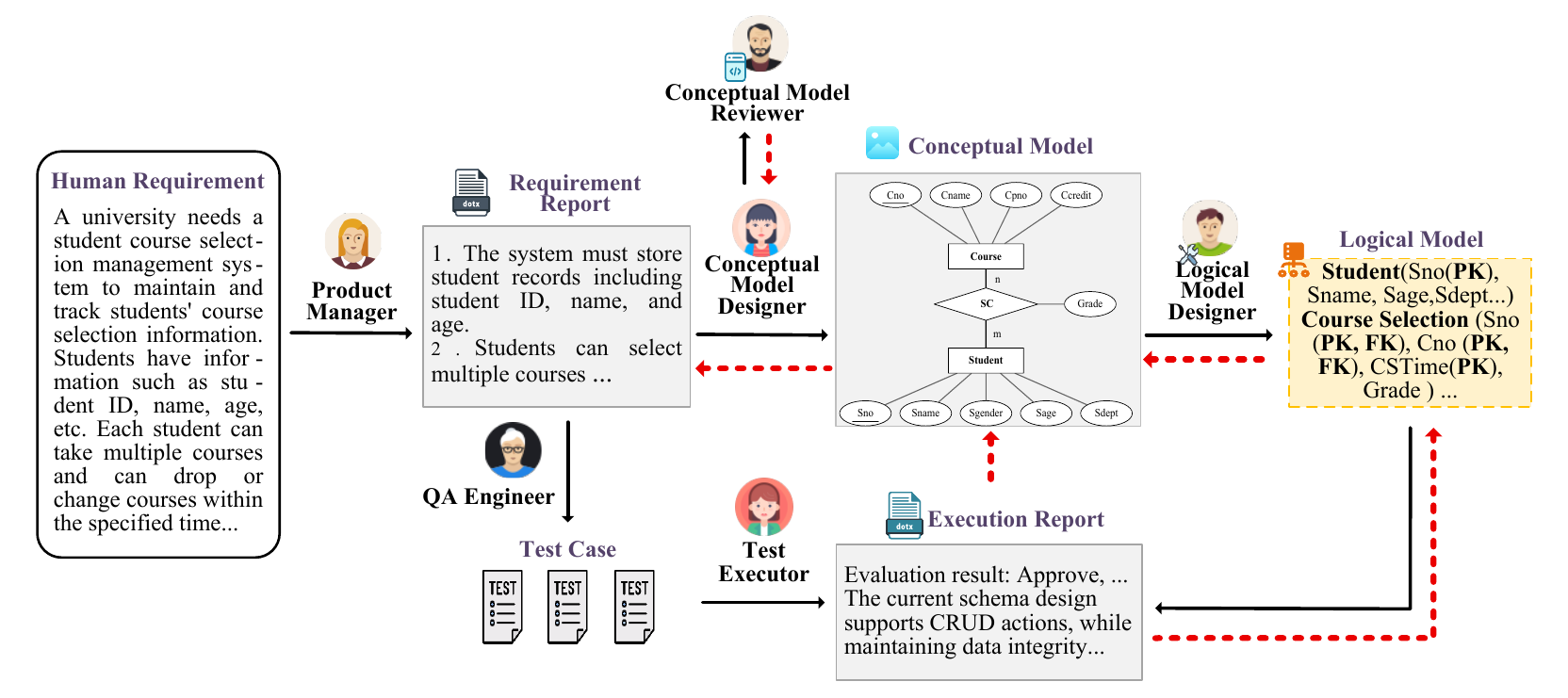}
	\caption {Our \textit{SchemaAgent} framework uses LLM-based agents with six distinct roles, including Project manager, Conceptual model designer, Conceptual model reviewer, Logical model designer, QA engineer, and Test executor. They collaboratively handle the sub-tasks involved in designing the database schema. The red arrow represents the process of error detection and correction.} 
\label{pic:framework}
\end{figure*}

We divide the standard database schema generation process into several subtasks, each of which is managed by a specialized agent, as indicated in Figure~\ref{pic:framework}.
There are in total six roles in our SchemaAgent framework. 
These roles collaboratively adhere to the systematic database design workflow to generate a comprehensive database schema. 
Each LLM-based agent in SchemaAgent operates according to a meticulously crafted profile that includes job description, goal, constraints, specialized knowledge and output format. All agents adhere to the React-style behavior as detailed in~\cite{yao2023react}. The profiles of each role, along with their associated tasks, are presented below.

\textbf{Product Manager (PM) agent} is the role that interacts directly with the user. It is mainly responsible for conducting the requirement analysis and generating the functional requirement analysis report. As illustrated in Listing~\ref{lst:product_manager}, we provide a static example as a demonstration. 
This demonstration provides end-to-end coverage of our system workflows, with each step’s output acting as a template for the respective agent. This standardization promotes uniform formatting and improves the agent's comprehension of tasks.
% The product manager agent accepts user requirements as input and produces a detailed requirements analysis report. 
\begin{lstlisting}[style=mystyle, caption={Format of Product Manager agent}, label={lst:product_manager}]
You are an experienced product manager.
# Goal: 
Generate requirement analysis reports: You are responsible for analyzing user requirements and clarifying any ambiguities by incorporating real-world scenarios, ensuring that the requirements are clearly defined ...
# Example: {example}
# Input: {input}
\end{lstlisting}

\textbf{Conceptual Model Designer (CMD) agent} identifies the components (i.e., the entity set, relationship set, mapping cardinality and attributes of the entity/relation set) of the conceptual model. 
Listing~\ref{lst:conceputal_model_designer} specifies constraints for the conceptual model designer, mandating a clear separation between entity sets and relationship sets, as their roles and operational behaviors in the data model are fundamentally distinct. Additionally, relationship sets typically employ composite keys rather than IDs to enforce database normalization and avoid redundancy.
\begin{lstlisting}[style=mystyle, caption={Format of Conceptual Model Designer agent}, label={lst:conceputal_model_designer}]
You are an expert in building database entity-relationship models.
# Goal: Based on the requirements analysis report, define the entity sets ... to build a database entity-relationship model.
# kownledge: 
  - An entity is a "thing" or "object" ... 
  - A relationship is a mutual association between multiple entities ... 
  - The mapping cardinality represents the number of other entities ...
# Constraint: 
  - Entity set names are mostly nouns, and relationship set names are mostly verb or verb-object structures ... 
  - Most relationship set attributes should not contain IDs. 
  ...
# Output Format:
  - If you have any uncertainties when identifying entities, ... send the issue to the ManagerAgent. If you have no questions, the conceptual model design is filled in "output". 
  - Your final answer is the JSON format converted from the entity-relationship model.
# Example: {example}
# Input: {input}
\end{lstlisting}
% - Entity sets (rectangles) are connected to relationship sets (diamonds).
% - Convert relationship sets to binary relationship sets when appropriate.

\textbf{Conceptual Model Reviewer (CMR) agent} provides essential and timely feedback on the conceptual model, using pseudocode-styled prompts to ensure thorough evaluation. As the conceptual model stands as the core of the entire process, a meticulous review is critical. Listing~\ref{lst:conceputal_model_reviewer} outlines pseudocode-based validation for entity/relationship sets, verifying: (1) absence of redundant IDs in relationship sets, (2) valid mapping cardinalities, and (3) correct entity references. 
Upon identifying errors, the agent returns detected errors with remediation suggestions to the conceptual model designer agent, triggering the regeneration of an improved conceptual model.
\begin{lstlisting}[style=mystyle, caption={Format of Conceptual Model Reviewer agent}, label={lst:conceputal_model_reviewer}]
You are a reviewer of the conceptual model of a database.
# Goal: You will judge whether the conceptual model meets its constraints.
# Knowledge: For the conceptual model, you have some evaluation criteria described in the form of pseudocode. The pseudocode is as follows:
 ```python
   FUNCTION ValidateData(json_data):
     entity_sets = json_data["output"]["Entity Set"]
     relationship_sets = json_data["output"]["Relationship Set"]
     # Step 1: Validate Relationship Set
     FOR relationship_name, relationship_details IN relationship_sets:
     # 1.1 Check if relationship attributes do not contain IDs
     IF Contains_ID_Without_Use(relationship_details["Relationship Attribute"]):
       logger.info "Relationship set " + relationship_name + " is not standardized: Attributes should not contain IDs."
     ...      
     logger.info "Validation completed."
    ```
# Output Format:
  (1) If the conceptual design does not meet these constraints, please send your suggestions to ConceptualDesignerAgent.
  ...
# Example: {example}
# Input: {input}
\end{lstlisting}

\textbf{Logical Model Designer (LMD) agent} transforms the conceptual model into a normalized logical schema by identifying functional dependencies and data types. As illustrated in Listing~\ref{lst:logical_model_designer}, this agent may employ our encapsulated tools for primary key identification and schema decomposition based on the Armstrong axioms~\cite{armstrong1974dependency} and the closure computation rules. This process ensures alignment with the 3NF normalization standards.
We adopt a coarse-grained approach to data types, limiting categories to NUMERIC, TEXT, DATETIME, BINARY, and BOOL, without distinguishing finer distinctions like BIGINT or TINYINT. Given that the requirement description currently focuses on functional requirements, \textit{not null} constraint and \textit{unique} constraint have been only applied to primary keys. Further constraints will be implemented as more detailed and explicit requirements become available.
\begin{lstlisting}[style=mystyle, caption={Format of Logical Model Designer agent}, label={lst:logical_model_designer}]
You are an expert in building the logical model of a database.
# Goal: Obtain a database relational schema that conforms to the third normal form based on the conceptual design of the database. 
# Knowledge: {knowledge}
# Constraint: 
- Identify functional dependencies and data type in all entity sets.
- Use the provided tool to identify the primary keys of all entity sets in the conceptual model. If any entity set lacks a primary key, the conceptual design is deemed invalid. Abort the task and report the error to the ConceptualDesignerAgent.
...
# Output format: 
- If you find any errors during task execution, you need to fill in these errors ...
# Example: {example}
# Input: {input}
\end{lstlisting}

\textbf{QA Engineer (QAE) agent} generates natural language test cases based on the requirement analysis report. As illustrated in Listing~\ref{lst:qa_engineer}, these cases simulate real-world operations such as insertion, deletion, update, and query, with actual data values. 
\begin{lstlisting}[style=mystyle, caption={Format of QA Engineer agent}, label={lst:qa_engineer}]
You are a quality assurance expert in database design.
# Goal:
According to the requirements analysis, you will generate 10 sets of test data, each of which includes specific values for four operations: insert, delete, query, and update.
# Knowledge: {knowledge}
# Constraint: 
Your test cases must consider aspects such as entity integrity, referential integrity, etc.
# Example: {example}
# Input: {input}
\end{lstlisting}

\textbf{Test Executor (TE) agent} could understand the test cases generated by the QAE and evaluate if the designed schema meets the testing criteria, culminating in a comprehensive test report.
\begin{lstlisting}[style=mystyle, caption={Format of Test Executor agent}, label={lst:test_executor}]
You are a database expert. 
# Gold: You can understand database operations described in natural language and judge whether the current schemas can meet the operational requirements. 
# Constraint: 
- If the current schema cannot pass your test, the design is unreasonable, and you need to send the error report to the role in charge who can solve the problem ...
- If you think it is reasonable after testing, fill with "TERMINAL" ...              
# Example: {example}
# Input: {input}
\end{lstlisting}

Following the comprehensive process involving requirement analysis, conceptual modeling, logical schema transformation and normalization, and rigorous testing, SchemaAgent ultimately produces a detailed relational database logical schema.

\subsection{Group Chat Communication Mechanism}
We implement the workflow using a group chat communication mechanism~\cite{wu2023autogen, hong2023metagpt}, where agents (speakers) communicate with each other through a shared message pool, managed by a group administrator that is essentially an LLM.
This mechanism could achieve an agent interaction that is more efficient than peer-to-peer interactions. 
More importantly, we establish an additional nested group over the CMD and the CMR to facilitate their closer communication. 
In this way, discussions related to conceptual model design would be confined within this nested group and transparent to other agents. 

By default, our system follows a fixed workflow, as illustrated by the black arrows in Figure~\ref{pic:framework}. 
In addition to the user input and requirement report, which are globally accessible to all agents, each agent maintains its own context consisting of all messages it has sent and received. The outputs produced by the agents are then deposited in the shared message pool.

% We incorporate specific instructions into agents' profiles for error feedback. Upon error feedback happening, the agent will generate an identifier for the error-handling agent within its output. This mechanism enables us to precisely determine the subsequent speaker in the communication flow.

\subsection{Controllable Error Detection and Correction}
We propose a controllable mechanism for error detection and correction.
Currently, there are two drawbacks in our framework.
Firstly, the sequential workflow tends to accumulate errors over time. 
% Secondly, the order of speakers can get out of control in group chat, which may bring a negative impact to the workflow.
Secondly, fixed speaking order hinders the effective transmission of error feedback.
In fact, we find that there could be a certain delay in error detection.
And we intend to enable each agent to help identify possible errors from previous ones in the workflow, so that we could reduce the accumulated errors.

\begin{figure*}[!t]
	\centering
	\includegraphics[width=0.9\linewidth]{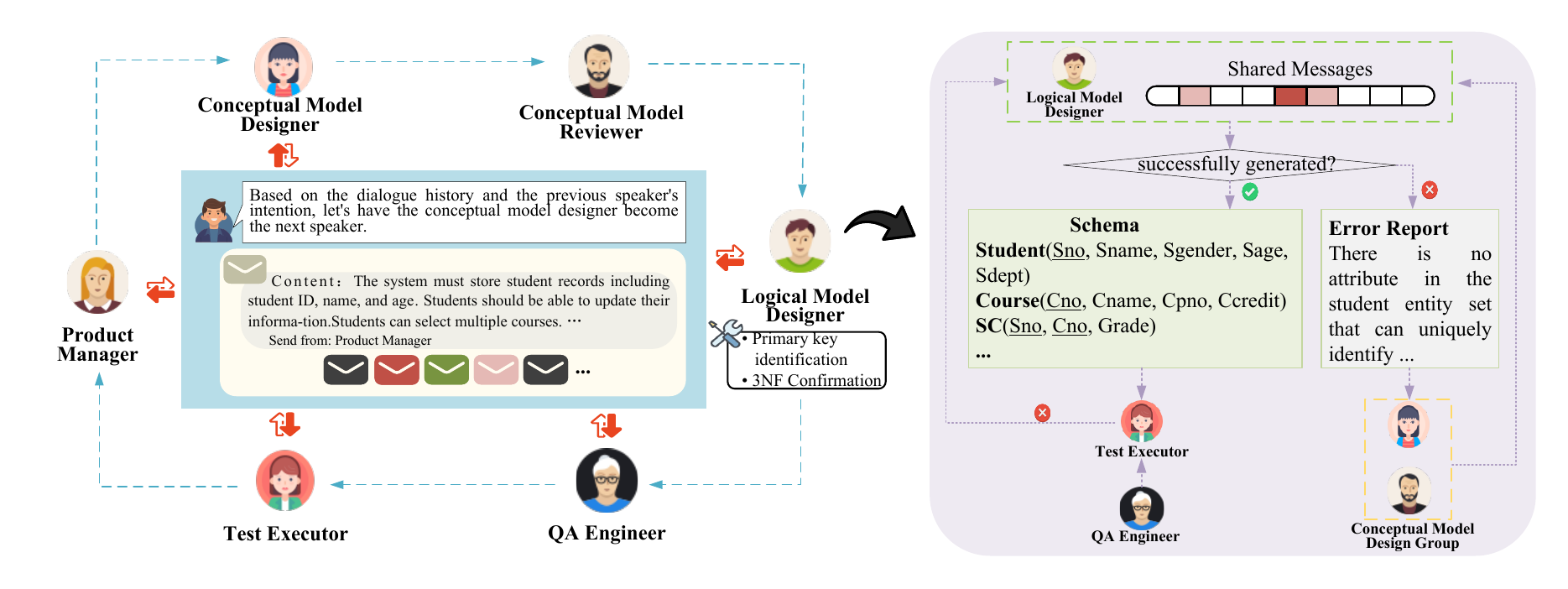}
	\caption {The error feedback process of the logical model designer in our group chat communication mechanism.} 
\label{pic:feedback}
\end{figure*}

In this way, we design a mechanism to control the speaker order in the group.
Specifically, for each role, we select a set of other roles that could be candidate next speakers. 
The arrows in Figure~\ref{pic:framework} indicate such candidate relations, where the black ones constitute the sequential part of the workflow, while the red dotted ones are for error feedback.
For example, at the time when the LMD outputs a schema, it may detect errors in the conceptual model, and hence the CMD could be a possible next speaker. Also, if there is no error, the schema could be an input to generate the execution report by the TE, which could similarly be the next speaker. Overall, there are two next candidate speakers, that is, CMD and TE, after the LMD speaks (generating schema). Figure~\ref{pic:feedback} presents an example of the LMD error feedback process. Upon identifying functional dependencies, if the LMD finds that there are entities or relations with no primary keys, it sends the error message back to the CMD, requesting a re-generation of the conceptual model. 

% During implementation, we integrate specific instructions into agents' profiles for error feedback. 
% The agent generates an identifier in its output, designating the appropriate error-handling agent from its list of candidates.
% This identifier allows for the determination of the subsequent speaker.
We can implement this by integrating these candidate relations into the agent's profile, so that the agent that is speaking can dynamically determine the next speaker from the candidate ones accordingly. 
Specifically, the agent generates an identifier in its output, designating the appropriate agent from its list of candidates.
This identifier allows for the determination of the subsequent speaker.
In cases where no candidate is selected, a predefined forward speaking order serves as a fallback mechanism to determine the next speaker.
This mechanism ensures that the conversation remains coherent and progresses efficiently, while allowing for flexible and context-aware decision-making in multi-agent interactions.

\section{Settings}
\subsection{Dataset}
We utilize the \textit{Rschema} dataset introduced in this paper, which comprises carefully constructed database design instances covering multiple domains, to effectively assess the performance of models in practical database design tasks. 

\subsection{Implementation Details} 
In all of our experiments, we use OpenAI's API services. The temperature and top p are set to 1.0. The group chat session is terminated upon the receipt of a message containing the keyword 'TERMINAL'. Given the inherent complexity of the task, there is a theoretical possibility that certain instances may fail to converge. To prevent infinite loops in such cases, we set an upper bound of 15 interaction rounds. Furthermore, while we implement various JSON parsing mechanisms, we configure a retry limit of three attempts to prevent unforeseen errors. This measure ensures that all test cases yield a parsable result. Statistically, SchemaAgent takes an average of 35 seconds and \$0.05 to generate a single case. 

\subsection{Evaluation Metrics}
% While we acknowledge that multiple valid schemas may exist to satisfy user requirements,
Manually evaluating each schema is prohibitively time-consuming and makes it difficult to quantitatively assess model performance. Consequently, we introduce an automated evaluation approach, using the manually annotated schemas in RSchema as the ground truth. These schemas undergo multiple rounds of evaluation, which ensures their correctness and superiority.
As mentioned in Section ~\ref{subsec:task_definition}, a schema consists of five key components: relation schema(table), attribute, primary key, foreign key, and constraint.
We measure the average F1 score and exact matching accuracy (Acc.) between the predicted and ground truth values for these four components. Acc. is set to 1 only if the F1 score equals 1; otherwise, Acc. is set to 0. 

\textbf{Relation schema (Table)} Since generative models may produce different tokens with similar meanings, we employ three alignment methods to match predicted schema names with ground truth counterparts. 
(1) Synonym Matching: WordNet~\cite{miller1995wordnet} is used to extract synonyms of predicted values and verify the inclusion of the ground-truth value within this set. 
(2) Similarity Matching: all-MiniLM-L6-v2~\cite{reimers2019sentence} is utilized to measure the semantic similarity between predicted and ground-truth values. The threshold $\delta_0$ is set to 0.6. 
(3) String Matching: We calculate the longest common substring between the predicted and ground-truth values and check if its length exceeds the threshold $\delta_1$ (set to 0.75). 

There are many tables in a schema. The formula~\ref{equ1} illustrates the computation of table F1 for each schema.
\begin{equation}
        \text{F1}_{table} = 2 \times \frac{\text{P} \times \text{R}}{\text{P} + \text{R}}
	\label{equ1}        
\end{equation}
\begin{equation}
    \text{P} = \frac{|\mathcal{R}_{gt} \cap \mathcal{R}_{pred}|}{|\mathcal{R}_{pred}|}, 
    \text{R} = \frac{|\mathcal{R}_{gt} \cap \mathcal{R}_{pred}|}{|\mathcal{R}_{gt}|}
\end{equation}
Where $\mathcal{R}_{gt}$ refers to the set of relation schemas in the ground truth schema, and $\mathcal{R}_{pred}$ refers to the set of relation schemas in the predicted schema.

\textbf{Attribute} We also apply the three alignment methods mentioned above to calculate the F1 score between the golden attribute set and the predicted attribute set in the mapped relation schema. Acc. is set to 1 only if F1 equals 1. For attributes in the golden tables that lack a corresponding mapping to the predicted tables, all metrics for those attributes are set to 0. Since a schema contains many attributes that belong to different relation schemas. The calculation method for attribute F1 of each sample is detailed in formula~\ref{equ2}.
\begin{equation}
    \text{F1}_{attrs} = \frac{1}{|\mathcal{R}_{gt}|} \sum_{\mathcal{R}_{gt,i} \in \mathcal{R}_{gt}} \text{F1}_{attrs}(\mathcal{R}_{gt,i})
    \label{equ2}
\end{equation}
Where $\text{F1}_{attrs}(\mathcal{R}_{gt,i})$ is computed by comparing the attribute set of the gold schema $\mathcal{R}_{gt,i}$ with that of the predicted one.
% $\text{Precision}_{attrs}(R_{gt,i})$ is the proportion of correctly predicted attributes in the predicted relation schema. $\text{Recall}_{attrs}(R_{gt,i})$ is the proportion of golden attributes that were correctly predicted. If a golden relation schema has no corresponding mapped schema in the prediction, its $\text{F1}_{attrs}(R_{gt,i})$ is set to 0.

\textbf{Key} Unlike relation schema and attributes, we advocate complete matching of primary keys and foreign keys. Acc. is 1 only when the golden key set and the predicted key set are fully identical.

\textbf{Data type \& Constraint} It depends solely on attributes with minimal impact on the overall schema structure. We only evaluate data type Acc. for successfully matched attributes, as unmatched attributes yield a score of zero. 
Due to the fact that \textit{not null} constraints and \textit{unique} constraints are only on the primary key, we have omitted the evaluation of constraints.

\begin{figure}[!htpb]
	\centering
	\includegraphics[width=3.0in]{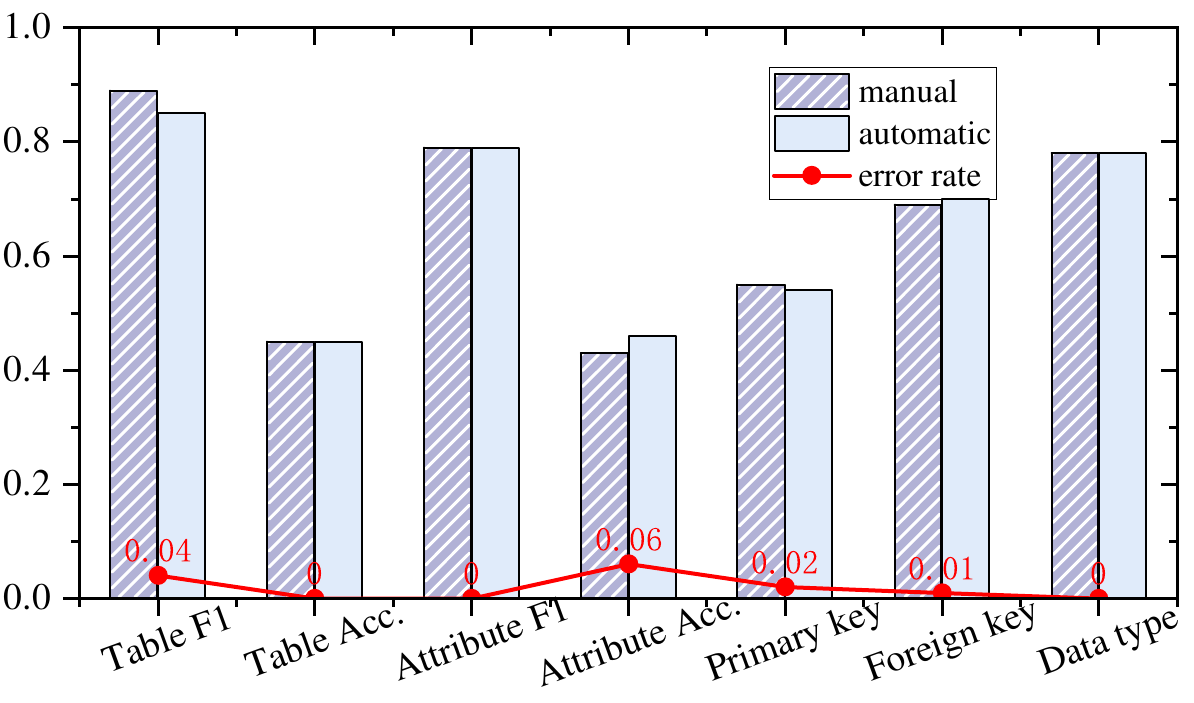}
        \caption{The results of manual evaluation and automated evaluation, as well as the error rate of the automated evaluation.}
\label{pic:error_rate}  
\end{figure}

These metrics are primarily based on semantic matching and may inevitably include some errors. 
We perform a manual evaluation of 20 randomly selected cases that includes more than 100 tables, and the results are presented in Figure~\ref{pic:error_rate}. As depicted in the figure, the discrepancy between the two methods is not statistically significant. The average error rate of our designed evaluation metric is 1. 8\%, which is within an acceptable range. The comparison with manual evaluation results proves that our evaluation system is generally reliable and effectively reflects the quality of the model's performance. 

\subsection{Baseline}
In the SchemaAgent framework, all agent capabilities are powered by the same LLM, such as GPT-3.5-turbo, GPT-4o, and DeepSeek-v3. In the one-shot setting, a case will run for all samples. 
In the few-shot setting, we additionally include a simple case with only two tables, a moderate case with four tables, and a complex case with six tables to facilitate context learning. In the CoT setting, we have manually defined six crucial steps for schema construction, namely: (1) Identify core entities and their attributes from requirements; (2) Define relationships between entities and their cardinality; (3) Map conceptual model to relational schemas and keys; (4) Define attribute domains and integrity constraints; (5) Normalize the database to avoid redundancy; (6) Consider other necessary elements. 

\section{Experimental Results and Analysis}
\label{Experimental Results and Analysis}
\subsection{Main Result}
Table \ref{table:main} presents the results on the RSchema benchmark dataset. These metrics reflect whether the design meets user requirements. We compare SchemaAgent with several mainstream baseline models in CoT, one-shot, and few-shot settings. We can observe from the experimental results that:

\begin{table}[htpb]
    \centering
    \caption{Experimental results of several methods on RSchema dataset. PR. refers to primary key. FK. refers to foreign key and DT. refers to data type.}
    \small
    \setlength{\tabcolsep}{0.15cm} %设置表格宽度
    \begin{tabular}{lccccccc}
        \toprule
        \multirow{2}*{\textbf{Method}} & \multicolumn{2}{c}{\textbf{Table}} & \multicolumn{2}{c}{\textbf{Attribute}} & \textbf{PK.} & \textbf{FK.} & \textbf{DT.} \\
        \cmidrule(lr){2-3}\cmidrule(lr){4-5}\cmidrule(lr){6-6}\cmidrule(lr){7-7}\cmidrule(lr){8-8}
        & F1 &Acc. &F1 &Acc. &Acc. &Acc. &Acc. \\
        \cmidrule(lr){1-8}
        \multicolumn{8}{c}{\textbf{GPT-3.5-turbo}} \\        
        One-shot &86.09 &51.97 &71.65 &32.81 &53.02 &74.01 &81.08 \\
        One-shot+CoT &86.09 &51.71 &70.87 &33.60 &52.49 &74.54 &81.34 \\
        Few-shot &85.56 &49.34 &71.13 &34.38 &54.07 &74.80  &81.72 \\
        SchemaAgent &\textbf{89.50} &\textbf{61.94} &\textbf{77.69} &\textbf{44.09} &\textbf{64.57} &\textbf{79.27} &\textbf{82.50} \\
        \cmidrule(lr){1-8}
        \multicolumn{8}{c}{\textbf{GPT-4o}} \\
        One-shot &85.30 &51.18 &69.82 &35.70 &55.12 &70.34 &82.06 \\
        One-shot+CoT &87.14 &54.59 &71.13 &36.48 &56.17 &71.92 &81.94 \\
        Few-shot &88.71 &58.79 &71.92 &35.96 &57.74 &71.92 &82.96 \\
        SchemaAgent &\textbf{90.29} &\textbf{65.09} &\textbf{79.53} &\textbf{49.87} &\textbf{73.23} &\textbf{81.63} &\textbf{83.76} \\
        \cmidrule(lr){1-8}
        \multicolumn{8}{c}{\textbf{DeepSeek-v3}} \\
        One-shot &80.58 &46.72 &67.19 &32.55 &46.72 &63.52 &80.08 \\
        One-shot+CoT &85.30 &49.61 &67.98 &35.96 &46.46 &62.20 &81.15 \\
        Few-shot &86.88 &53.81 &72.70 &38.06 &51.18 &62.73 &81.60 \\
        SchemaAgent &\textbf{88.98} &\textbf{61.68} &\textbf{78.48} &\textbf{46.72} &\textbf{71.92} &\textbf{80.84} &\textbf{82.35} \\
 
        \bottomrule
    \end{tabular}
    \label{table:main}
\end{table}

\textbf{SchemaAgent outperforms other prompt-based baselines.} Our proposed framework significantly outperforms the corresponding baseline models mostly across all metrics, demonstrating that our framework could generate higher-quality database logical schemas.

\textbf{CoT offers modest gains.} Introducing a reasoning step via CoT provides a slight performance boost compared to the one-shot method. This suggests that prompting the model to "think step-by-step" aids in the nuanced task of identifying schema components.

\textbf{Few-shot outperforms the CoT strategy in most metrics.} The few-shot prompting method provides concrete examples of desired outputs and their corresponding inputs. Few-shot learning leads to a more refined understanding of the task, enabling the model to generate more accurate schemas across a broader range of use cases, thereby consistently outperforming approaches solely relying on the CoT strategy across various evaluation metrics in the schema generation task.

\textbf{The backbone model is important.} In the LLM-based agent framework, GPT-4o generally outperforms DeepSeek-v3 and GPT-3.5-turbo, highlighting the importance of the foundational capabilities of the agents.

\begin{table*}[hptb]
    \centering
    \caption{The contribution of roles. The avatar icons in the "Roles" column from left to right represent "Product manager", "Conceptual model designer", "Conceptual model reviewer", "Logical model designer", "QA engineer" and "Test executor".}
    \small
    \setlength{\tabcolsep}{0.3cm} %设置表格宽度
    \begin{tabular}{cccccccccccc}
        \toprule
        \multicolumn{6}{c}{\textbf{Roles}}
        & \multicolumn{2}{c}{\textbf{Table}} & \multicolumn{2}{c}{\textbf{Attribute}} & \textbf{Primary key} & \textbf{Foreign key} \\
        \cmidrule(lr){1-6}\cmidrule(lr){7-8}\cmidrule(lr){9-10}\cmidrule(lr){11-11}\cmidrule(lr){12-12}
        \includegraphics[width=6mm]{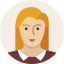} &\includegraphics[width=6mm]{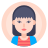}
        &\includegraphics[width=6mm]{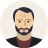}
        &\includegraphics[width=6mm]{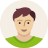}
        &\includegraphics[width=6mm]{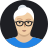}
        &\includegraphics[width=6mm]{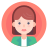}
        &\raisebox{1ex}{F1} &\raisebox{1ex}{Acc.} &\raisebox{1ex}{F1} &\raisebox{1ex}{Acc.} &\raisebox{1ex}{Acc.} &\raisebox{1ex}{Acc.}\\
        \cmidrule(lr){1-12}
        \includegraphics[width=4mm]{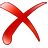} &\includegraphics[width=4mm]{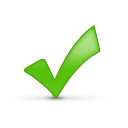} &\includegraphics[width=4mm]{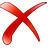} &\includegraphics[width=4mm]{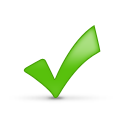} &\includegraphics[width=4mm]{latex/no.png} &\includegraphics[width=4mm]{latex/no.png}
        &87.14 &54.86 &70.60 &40.68 &66.40 &67.98 \\
        
        \includegraphics[width=4mm]{latex/yes.png} &\includegraphics[width=4mm]{latex/yes.png} &\includegraphics[width=4mm]{latex/yes.png} &\includegraphics[width=4mm]{latex/yes.png} &\includegraphics[width=4mm]{latex/no.png} &\includegraphics[width=4mm]{latex/no.png} 
        &88.71 &61.15 &74.80 &44.09 &70.87 &80.58 \\  
        
        \includegraphics[width=4mm]{latex/yes.png} &\includegraphics[width=4mm]{latex/yes.png} &\includegraphics[width=4mm]{latex/no.png} &\includegraphics[width=4mm]{latex/yes.png} &\includegraphics[width=4mm]{latex/yes.png} &\includegraphics[width=4mm]{latex/yes.png}
        &87.66 &59.06 &74.54 &39.37 &64.57 &78.22 \\     
       
        \centering\includegraphics[width=4mm]{latex/yes.png} &\includegraphics[width=4mm]{latex/yes.png} &\includegraphics[width=4mm]{latex/yes.png} &\includegraphics[width=4mm]{latex/yes.png} &\includegraphics[width=4mm]{latex/yes.png} &\includegraphics[width=4mm]{latex/yes.png} 
        &90.29 &65.09 &79.53 &49.87 &73.23 &81.63 \\
        
        \cmidrule(lr){1-12}
        \multicolumn{12}{c}{\textbf{w/o error detection and correction}} \\
        \centering\includegraphics[width=4mm]{latex/yes.png}
        &\includegraphics[width=4mm]{latex/yes.png} &\includegraphics[width=4mm]{latex/yes.png} &\includegraphics[width=4mm]{latex/yes.png} &\includegraphics[width=4mm]{latex/yes.png} &\includegraphics[width=4mm]{latex/yes.png} 
        &89.50 &62.20 &78.22 &49.08 &71.65 &81.10 \\
        
        \bottomrule
    \end{tabular}
    \label{table:roles}
\end{table*}

\subsection{The Number of Agent Roles}
We conduct an ablation study to better understand the contribution of each agent. 
The conceptual model designer and the logical model designer are two fundamental roles in our framework, which would not be removed in any case. We take the other four roles as variables to demonstrate their performance.
Upon removing one or more variable roles, we always keep the remaining agents to retain their original functionality to evaluate performance. 
As shown in Table~\ref{table:roles}, when all agents utilize GPT-4o, the inclusion of all variable roles significantly improves performance, indicating the importance of specialized roles in achieving optimal results.

Specifically, the conceptual model reviewer emerges as the most critical role, significantly improving all metrics. This indicates that the pseudocode-based validation prompts employed by the conceptual model reviewer effectively identify errors in the conceptual model and provide valuable feedback for their correction. The QA engineer and test executor play important roles in verifying the correctness of the logical model. Removing these two roles results in a performance decline. When only the fundamental conceptual model designer and logical model designer remain, the framework performs at its lowest level, struggling to accurately extract entities, relations, their associated attributes, and mapping cardinality.

\subsection{The Effectiveness of Agent Framework}
\begin{figure}[!htpb]
	\centering
	\includegraphics[width=\linewidth]{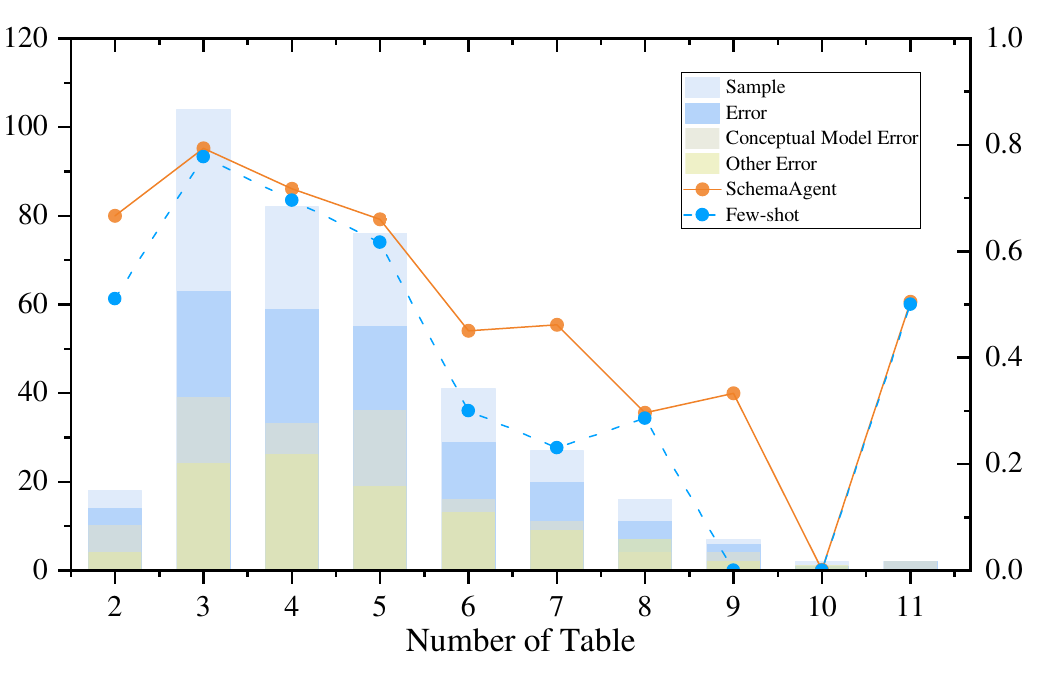}
	\caption {Error type distribution and SchemaAgent performance across varying table counts. The model performance is represented by the "Table Acc." score.} 
\label{pic:error_analysis}
\end{figure}

Our framework incorporates two error correction mechanisms. The first is a nested group, consisting of the CMD and CMR, dedicated to detecting and optimizing the conceptual model. The second mechanism involves error detection and correction among the other agents. Statistically, 69.9\% of samples receive error feedback. Of these, 60.0\% have errors exclusively identified and handled between the CMD and the CMR. Figure~\ref{pic:error_analysis} illustrates the frequency of each type of error in varying difficulty levels in the task, categorized by the number of tables. It also compares the performance of SchemaAgent against the suboptimal result (prompt with few-shot). As the figure shows, the frequency of error feedback steadily increases with the number of tables. In particular,\textit{ other errors} also account for a growing proportion of all errors, indicating a higher probability of latent errors emerging as task complexity increases. The improvement over the suboptimal result further demonstrates the superiority of our proposed SchemaAgent.

\subsection{The Effectiveness of Controllable Error Detection and Correction Mechanism}
To evaluate the proposed error detection and correction mechanism, we established a baseline model. This model follows a conventional schema generation process and includes no error feedback, except for the feedback in the nested group during conceptual model design.
In this baseline, each agent performs its designated task to the best of its ability and then passes the output to the next agent in a predetermined workflow. As illustrated in Table~\ref{table:roles}, SchemaAgent exhibits performance improvement after the utilization of the controllable error feedback and correction mechanism, with all metrics demonstrating an upward trend. 

\begin{figure}[!htpb]
	\centering
	\includegraphics[width=3.0in]{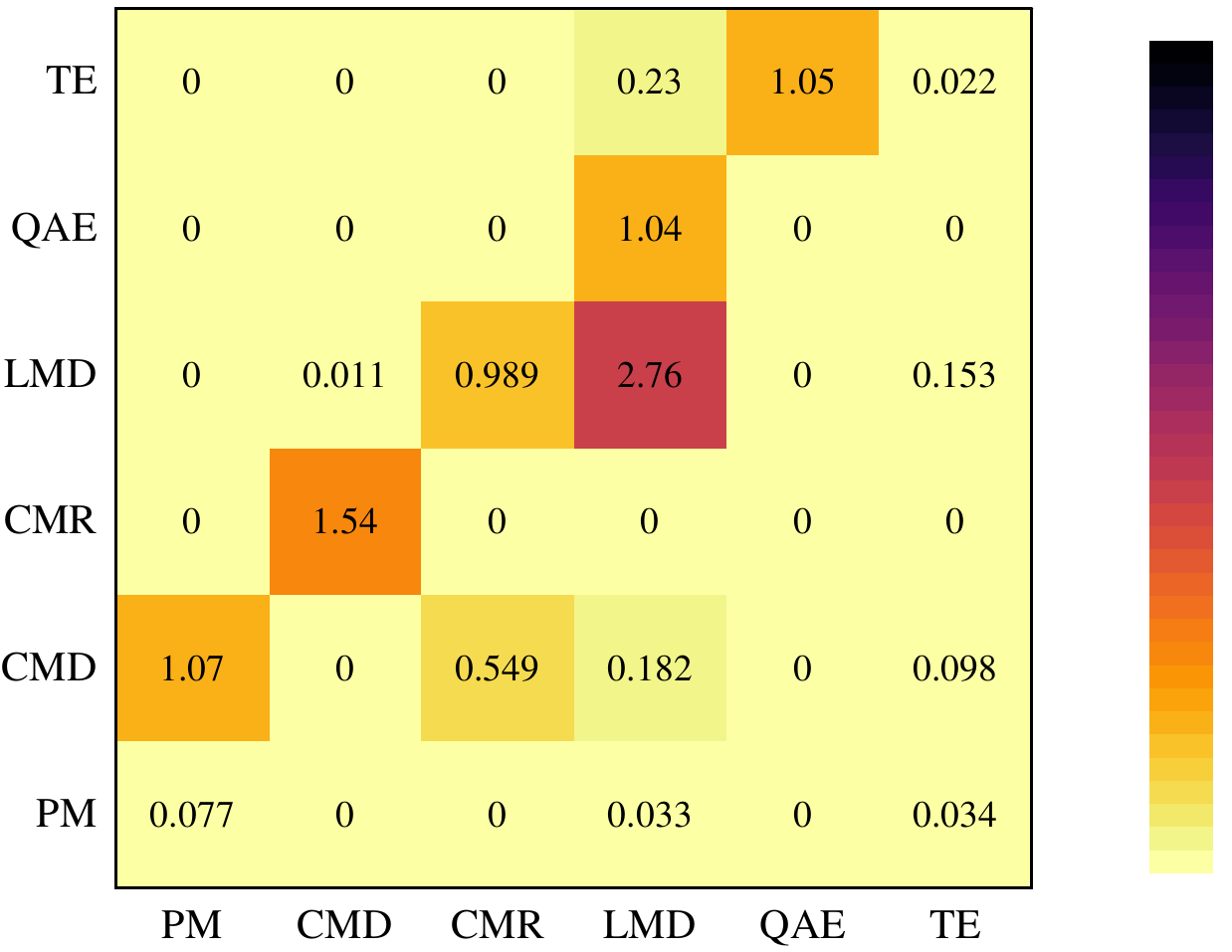}
	\caption {The communication frequency statistics of all agents in  the SchemaAgent framework. Messages are transmitted from roles located on the horizontal axis to roles located on the vertical axis.}     
\label{pic:feedback_count} 
\end{figure}

We further investigate the feedback process by analyzing the frequency of feedback interactions among agents. As depicted in Figure~\ref{pic:feedback_count}, we can observe that:
(1) Communication between the CMD and the CMR is the most frequent. This indicates that the CMR challenges the current conceptual model design in roughly half of the cases. Although its feedback may not always be entirely justified, this interaction facilitates a crucial design refinement process. This also highlights the critical and complex nature of the conceptual model design task. 
(2) The interaction frequency within the LMD itself is pretty high. This is due to the LMD making three additional calls to our encapsulated tools to generate a result: two tool calls for identifying the primary keys of entity and relationship sets, and one for lossless dependency decomposition. This frequency demonstrates that the LMD is able to leverage tools to solve problems in the vast majority of cases. 
(3) The CMD also receives error feedback from the LMD, whereas the LMD's feedback predominantly originates from the TE and PM. These feedback chains indicate that, under our controllable error detection and correction mechanism, the collaborative efficiency and accuracy of logical architecture generation have been improved.
(4) Some interactions are low-frequency due to the instability of LLMs; later-executing agents may only partially use the outputs (including their identifier) of previously running ones. This leads to unpredictable identifiers.

\subsection{LLM as The Judge}
Recognizing the inherent limitations of evaluating schema designs solely against a single, predefined ground truth, as multiple valid structures could satisfy user requirements in real-world scenarios, we use another LLM-based methodology for automated schema quality assessment. Specifically, \textit{deepseek-r1-0528} is employed as the adjudicator. We provide a granular evaluation of each generated schema across three critical dimensions, assigning a score on a 10-point scale (from 1 to 10) for each. These dimensions encompass: (1) Functional coverage, which assesses the completeness and extensibility of the schema in addressing business requirements; (2) Data redundancy and normalization, evaluating its adherence to normalization principles; and (3) Integrity constraints, examining the robustness of key and other relational rules. For each dimension, a score of 1 indicates fundamental deficiencies, while a score of 10 indicates exemplary adherence to best practices (i.e., 100\% functional coverage, absolute compliance with 3NF or comprehensive integrity implementation). 
\begin{figure}[!htpb]
	\centering
	\includegraphics[width=3.3in]{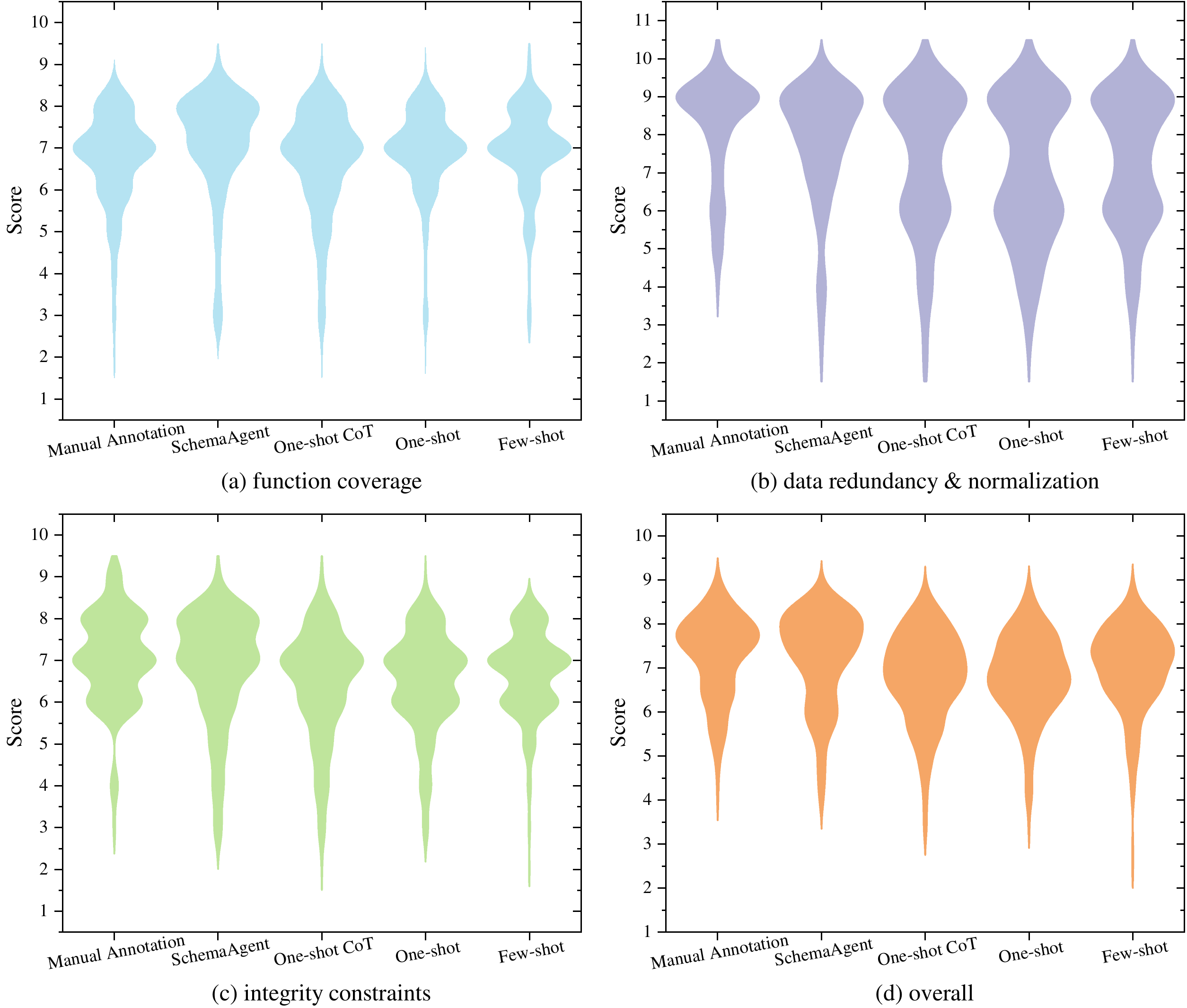}
	\caption {Evaluation of score distributions for different methods across various aspects in schemas.}     
\label{pic:llm_analysis}
\end{figure}

Figure~\ref{pic:llm_analysis} illustrates the score distributions. The overall score is a weighted average of function coverage, data redundancy \& normalization, and integrity constraints, with weights of 40\%, 30\%, and 30\%, respectively, reflecting the ease of user modification. Our analysis reveals that: (1) There is room for extensibility optimizations achieved by manual annotations, which are sometimes overlooked during manual process (e.g., setting an attribute instead of a table for expansion). (2) Overall, manual annotations perform strongly, validating the good quality of our dataset. In particular, scores for data redundancy and normalization are significantly higher for manual annotations, reflecting our design preference for consistency and elimination of redundancy. (3) SchemaAgent achieves the best results in all metrics in the baselines. Its strong performance in data redundancy and normalization further confirms our method's effectiveness in normal form control. (4) There exists some schemas from LLMs score higher than that from manual annotation, while, they account for a very small proportion, and we anticipate introducing extensibility optimizations in the future to address this issue.

\subsection{Performance Enhancement and Error Analysis}
\begin{figure}[!htpb]
	\centering
	\includegraphics[width=3.3in]{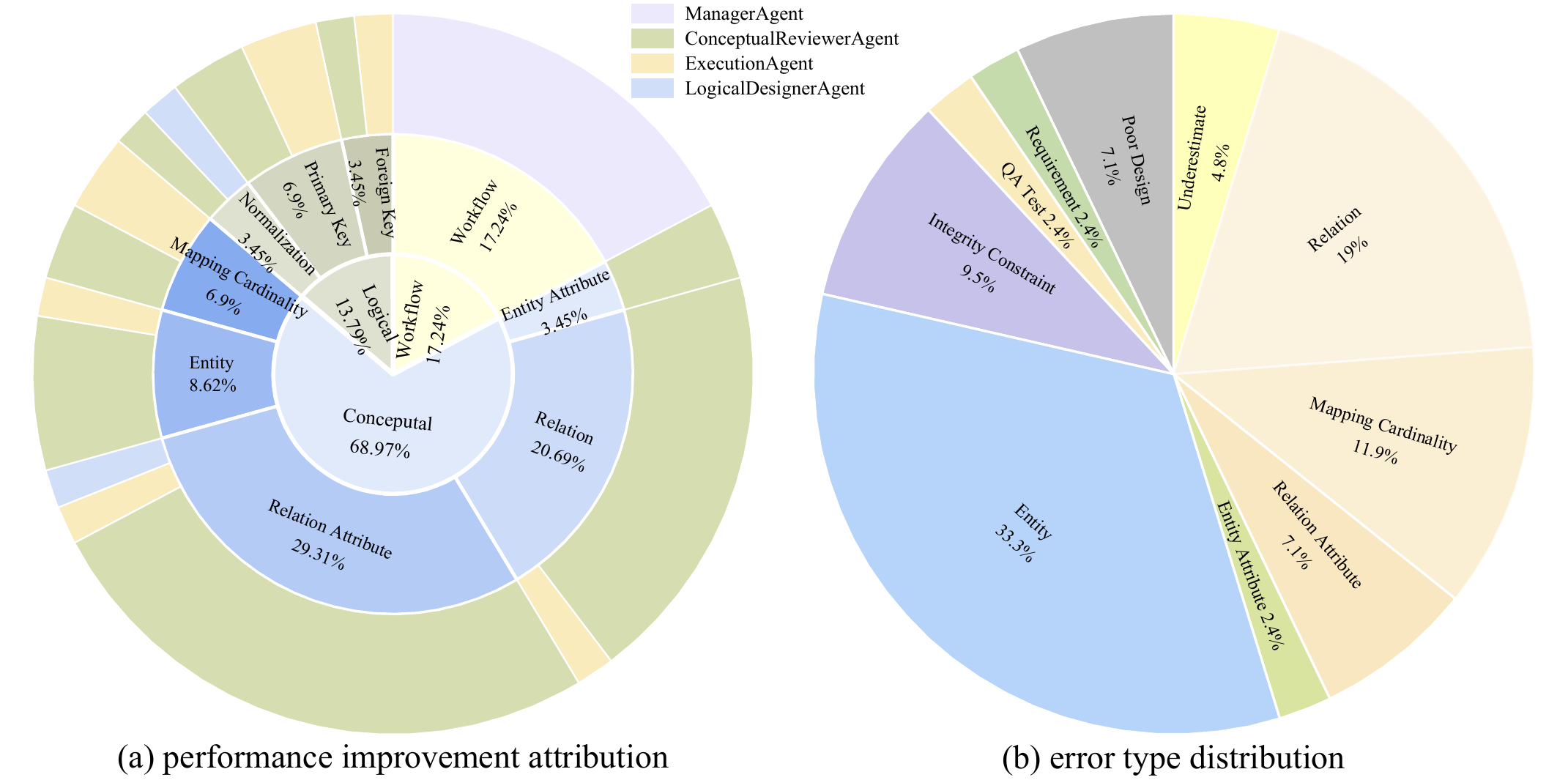}
	\caption {Performance improvement attribution and error type distribution.}     
\label{pic:performance_analysis}
\end{figure}

To evaluate how SchemaAgent improves performance, we analyze 31 high-scoring cases (based on automated metrics and LLM-as-the-judge). Performance improvements are categorized into three main areas: workflow (without error feedback), conceptual model, and logical model.
The conceptual model improvements involve enhanced recognition of entity, relation, and mapping cardinality. Logical model improvements cover primary/foreign key and normalization optimization. 
For each category, we trace the improvements back to the specific agents that provided feedback.
Figure~\ref{pic:performance_analysis}(a) shows 68.9\% of improvements come from the improvements of the conceptual model, with the CMR agent contributing 85\% of the feedback, and the LMD and TE agents provided the remaining ones.
In 13.79\% of the cases, with a correct conceptual model, LMD, TE, and CMR (and CMD) collaborate to optimize the logical model. 
When no feedback is required, SchemaAgent operates in a workflow-driven manner, directly generating high-scoring schemas. We attribute the improvement in this situation to the Product Manager agent. 

\begin{table*}[!t]
    \centering
    \caption{Comparison with Text2SQL models. "Executable" indicates the execution using SQLite in DDL format.}
    \small
    \setlength{\tabcolsep}{0.3cm} %设置表格宽度
    \begin{tabular}{lcccccccc}
        \toprule
        \multirow{2}*{\textbf{Method}} & \multicolumn{2}{c}{\textbf{Schema}} & \multicolumn{2}{c}{\textbf{Attribute}} & \textbf{Primary key} & \textbf{Foreign key} & \textbf{Data type} & \textbf{Executable}\\
        \cmidrule(lr){2-3}\cmidrule(lr){4-5}\cmidrule(lr){6-6}\cmidrule(lr){7-7}\cmidrule(lr){8-8}\cmidrule(lr){9-9}
        & F1 &Acc. &F1 &Acc. &Acc. &Acc. &Acc. &Acc.\\
        \cmidrule(lr){1-9}
        Arctic-Text2SQL-R1-7B &74.54 &42.26 &59.32 &33.33 &21.00 &38.06 &61.68 &91.86 \\
        OminiSQL-14B &84.25 &56.17 &69.55 &40.16 &63.78 &69.55 &61.61 &95.28 \\
        SchemaAgent &90.29 &65.09 &79.53 &49.87 &73.23 &81.63 &83.76 &97.14 \\
        \bottomrule
    \end{tabular}
    \label{table:ddl}
\end{table*}

To further evaluate SchemaAgent, we also analyze $31$ cases with low comprehensive scores, identifying ten error types. As shown in Figure~\ref{pic:performance_analysis}(b), conceptual design as a critical phase in the database design shows significant room for improvement. (1) Entity errors included both overidentifying entities (e.g., creating a separate \textit{recipient} table when recipients are already in the \textit{users} table) and underidentifying entities, especially when entities have brief descriptions. (2) Relation errors involve missing relationships between entities. (3) Relation attribute errors primarily concern adding redundant auto-incrementing ID attributes to relationships. (4) Mapping cardinality errors often misclassify many-to-many relationships (e.g., between \textit{supplier} and \textit{product}) as one-to-many, particularly when the requirement description is ambiguous, challenging models' understanding of real-world scenarios. (5) Due to the inherent limitations of automatic evaluation, an underestimate is observed in 4.8\% of the cases. (6) 7.1\% of functionally complete schemas lack extensibility, leading to poor design.
Furthermore, we observe that the model is more prone to errors when dealing with associative entities (entities with independent primary keys linked via foreign keys to other entities) and weak entities (entities whose existence depends on another entity). 
This suggests that the particular challenges in accurately discerning and representing complex entity relationships, especially those involving non-trivial primary key structures and existential dependencies, require advanced contextual understanding and external knowledge.

\subsection{Case Study for Schema Generation}
\begin{figure}[!htpb]
	\centering
	\includegraphics[width=3.3in]{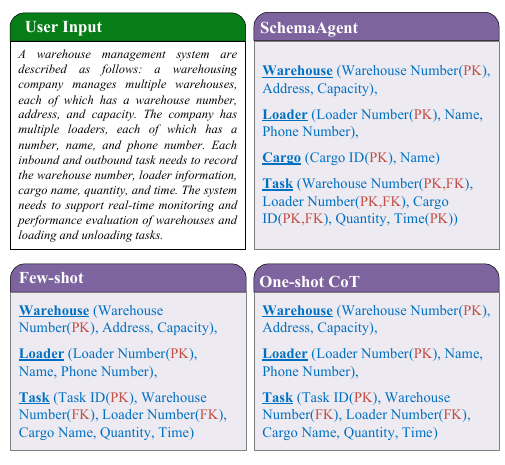}
	\caption {The schema output of SchemaAgent, few shot, and one shot with CoT in the warehouse management case.} 
\label{pic:case_study}
\end{figure}

To demonstrate the practical effectiveness of our proposed SchemaAgent framework, we present a case study in the domain of warehouse management. 
As shown in Figure~\ref{pic:case_study}, the schema generated under CoT and the few-shot prompt settings include the\textit{ Cargo Name} directly within the\textit{ Task} table, which introduces a risk of inconsistency in the data (e.g. spelling errors in the\textit{ Cargo Names}). 
The output of SchemaAgent adheres to the principles of the 3NF. In the \textit{Task} table, SchemaAgent correctly identifies the functional dependencies among attributes and accordingly determines a composite primary key. The normalized structure and composite key design significantly enhance data integrity. 

Our method demonstrates significant performance for generating normalized schemas. Future work could integrate user query patterns to facilitate strategic denormalization, which enables a balance between query efficiency and data consistency.

\subsection{Case Study for DDL Generation}
\label{sec:ddl_generation}
We demonstrate that it is easy to covert schemas into DDL, as a fundamental part of physical design.
Different database management systems (DBMS) employ diverse methodologies for physical design implementation.
To verify the validity of the schema, we use a rule-based approach to generate SQLite-specific DDL, which is subsequently executed to empirically verify our methodology. This treatment results in a process similar to the Text2SQL task. 
Therefore, we select prominent Text2SQL works for evaluation, including \textbf{Arctic-Text2SQL-R1}~\cite{yao2025arctic}: a reinforcement learning-based Text2SQL model that optimizes for generating executable SQL queries through execution feedback. \textbf{OmniSQL}~\cite{li2025omnisql}: a Text2SQL model trained on a large synthetic dataset. 

As shown in Table \ref{table:ddl}, existing Text2SQL models particularly struggle with relation schema identification. This highlights the crucial need for dedicated Text2DDL research, as it involves inferring complex structural design and relationships that existing Text2SQL approaches are not equipped to handle.

\section{Conclusion \& Future Work}
\label{sec:future_work}
The automation of database schema remains a significant challenge, due to its inherent dependence on specialized expertise and extensive accumulated practical experience.
In this paper, we propose, for the first time, an LLM-based multi-agent framework for logical schema generation. 
We call this framework SchemaAgent, where we design six roles with a controllable interaction mechanism for error detection and correction in the workflow.
In addition, we construct RSchema, a cross-domain dataset containing more than $381$ pairs of user requirement texts and their corresponding logical schemas. 
We systematically evaluate the performance of mainstream LLMs and SchemaAgent on RSchema. Experimental results demonstrate that SchemaAgent achieves state-of-the-art and highly competitive performance across all metrics. 

\textbf{Future Work.} 
This work is the first attempt to leverage LLMs for automated schema generation, which may give rise to several promising directions for future research, to name a few:
\begin{itemize}
    \item \textbf{Physical Design Integration.} 
    It would an interesting future work to incorporate physical design into our framework, which could make our work more comprehensive. Physical design may involve index generation, disk allocation, and so on.
    
    % \item \textbf{Mapping Cardinality Identification.} 
    % There have been many works on entity relation recognition, while, none of them consider mapping cardinality.
    % mapping cardinality defines the nature of relationships between entities and plays a crucial role in determining the appropriate logical schema structure. For example, many-to-many relationships require a separate join table, whereas one-to-one or one-to-many relationships may not. Unlike traditional tasks in natural language processing, such as entity recognition and relation extraction, cardinality types are rarely addressed, despite their importance in database design, making it a compelling area for further exploration.
    
    \item \textbf{Functional Dependency Discovery.} Identifying functional dependencies is a prerequisite for normalization. Traditional FDs discovery works are based on statistical analysis over datasets. However, with extensive world knowledge of LLMs, it is currently possible to discover FDs directly from natural language descriptions. Exploring LLM-based FD discovery opens up a new research avenue toward more intelligent and data-free schema normalization.
    
    \item \textbf{Query Efficiency and Normalization Balancing.} In this paper, we prioritize data consistency by requiring the generated schema to satisfy the 3NF. However, strict normalization may lead to suboptimal query performance in practice. A promising direction is to incorporate user query requirements into the schema generation process, aiming to achieve an optimal balance between data consistency and query efficiency.
\end{itemize}

\begin{acks}
We thank the following contributors for their assistance during the data annotation process: Chuchu Gao, Yuequn Dou, Zhaohai Sun, and Ke Tang.
\end{acks}

\balance  % 平衡最后一页的多栏内容

\clearpage

\bibliographystyle{ACM-Reference-Format}
\bibliography{sample}

\end{document}